%Paper: nucl-th/9412009
%From: moskowitz@sgs0.hirg.bnl.gov (Bruce Moskowitz)
%Date: Tue, 6 Dec 94 16:13:37 -0500

% begin TeX file
\magnification=\magstep1
\parskip=12pt
\baselineskip=18pt
\parindent 0.5truein

\centerline {\bf COMPARISON OF EXPERIMENTAL DATA TO THE}
\centerline {\bf RELATIVISTIC QUANTUM MOLECULAR DYNAMICS MODEL}
\centerline {\bf FOR Si+Au COLLISIONS AT 14.6 A GeV/c}

\vglue 0.5truein

\centerline {M.~GONIN$^a$, OLE HANSEN$^b$, B.~MOSKOWITZ, and F.~VIDEB\AE K}
\centerline {\it Department of Physics, Brookhaven National Laboratory,
Upton, NY 11973, U.S.A.}
\vglue 0.25truein
\centerline {H.~SORGE$^c$}
\centerline {\it Theory Division, Los Alamos National Laboratory,
Los Alamos, NM 87544, U.S.A.}
\vglue 0.25truein
\centerline {R.~MATTIELLO}
\centerline {\it Institut f\"ur Theoretische Physik,
Johann Wolfgang Goethe Universit\"at,}
\centerline {\it Frankfurt am Main, GERMANY}
\vglue 0.5truein
\leftline {\bf ABSTRACT}
\noindent
	Predictions from the RQMD model are systematically compared
	to recently published charged hadron distributions of AGS
	Experiment 802 for central Si+Au collisions at
	14.6$A$ GeV/$c$, taking into account both the experimental
	trigger condition and acceptance.
	The main features of the data, including K$^+$ production, can
be understood quantitatively to better than 20\% within the framework of
the model, although several discrepancies are found, most importantly
for the proton spectra.
\vglue 0.5truein
\leftline {\bf 1.~~INTRODUCTION}

The main purpose of heavy-ion experiments at high energies is to produce
and study nuclear matter under unusual conditions, e.g., at high
densities and temperatures. Several detailed models of nucleus-nucleus
collisions have been
developed (see, e.g., Refs. 1-4), all based on a fully confined phase
and on conventional particle production mechanisms.

The Relativistic Quantum Molecular Dynamics (RQMD) model$^2$
produces hadrons through the excitation of baryonic and
mesonic resonances.  Heavy resonances (more than 2 GeV for baryons and
more than 1 GeV for mesons)
are treated in the string picture following the Lund
model$^1$ and all particles are allowed to reinteract (baryon-baryon,
baryon-meson, and meson-meson).  The model provides a complete
time-dependent description of the evolution of each event.  The
probabilities for excitation of specific channels are governed by
experimental cross-sections to the extent possible.
The formation points of hadrons are taken from the properties of
resonance decay and string fragmentation.
Previous comparisons of RQMD predictions and AGS heavy ion data may
be found in Refs. 5-9.  In particular, the RQMD model gives a good
account of global observables, such as transverse energy$^5$ and
charged particle multiplicities$^6$.
Based on the successes with global observables,
it is expected that the model should
agree with the data at least at the 25\% level, where the
experimental systematic error for the data used here is 10-15\%.
This expectation is largely borne out by the present study.

The experimental data in this paper are from BNL-AGS Experiment
802 (E802), a movable single-arm magnetic spectrometer
setup$^{10}$.
Early E802 data from a run in December 1987 demonstrated a
K$^+/\pi^+$ ratio  of 19$\pm$1\% for 14.6$A$ GeV/$c$ central
Si+Au collisions$^{11,12}$, considerably above the p+p ratio of
5-6\% (see Ref.~13).  The data used in the present comparison are
from December 1988 and June 1989, are published in the survey paper
of Ref.~14, and encompass protons, pions, and kaons over
a broad rapidity range ($0.6 < y < 2.8$ for pions).
In this paper we concentrate our attention on the central Si+Au system.
What sets the present comparison aside from previous ones (see, e.g.,
Refs. 4,5) is a systematic approach starting from the trigger
conditions and encompassing all the data, including strangeness
production, stopping and momentum conservation.  Throughout
the paper the influence of the experimental acceptances on the model
results is given close attention.

\vfil\eject
\parskip=12pt
\parindent 0.5truein
\leftline {\bf 2.~~THE RQMD EVENT SAMPLES AND THE TRIGGER CONDITION}

The experimental trigger cross section $\sigma_{trig} = 263$ mb
corresponds to about 7\% of the total inelastic cross section for
Si+Au and is defined by a cut on ``apparent'' charged
multiplicities $>173$.  The word ``apparent'' stands for
multiplicities in the experimental acceptance, without corrections
for gamma conversion to e$^+$-e$^-$, multihits and delta
electrons.  The multiplicity condition was introduced in
software$^{14}$ and is different from the hardware trigger used
earlier$^{11,12}$.

The main RQMD sample consists of 1537 events generated by version
1.07 of the code, with impact parameters $0<b<4$ fm and a weighting
proportional to $b$, corresponding to a cross section of 503 mb
(see Fig.~1, solid histogram.)
For reference, the impact parameters $0<b<3.4$ fm correspond to a
full overlap between projectile and target nuclei in a hard sphere
picture.
To match the experimental $\sigma_{trig}$, 806 events of
the RQMD sample were
chosen by demanding that $n_{ch}\geq 154$, where
$n_{ch}$ is the charged multiplicity from an RQMD event in the ideal
experimental acceptance$^{10}$ ($6^\circ<\theta<149^\circ$,
where $\theta$ is the polar angle from the beam direction, and
momentum thresholds $p>0.30$ GeV/$c$ for protons and $p>0.05$ GeV/$c$
for pions).
The impact parameter distribution of this sample is shown by the
broken histogram in Fig.~1.  The $n_{ch}\geq 154$ sample
encompasses nearly all events with $b<2$ fm, but also has a
significant contribution from $2<b<3$ fm.
The sensitivity of the RQMD predictions to impact parameter selection
was tested and found to be small for events with $b<4$ fm.

The E802 cut at 7\% on apparent charged multiplicity does not
correspond to a sharp cut at 7\% in real charged multiplicity, as some
events with comparatively low charged particle multiplicity and a
comparatively high number of delta electrons and
gamma-conversion e$^+$e$^-$ pairs would be accepted by the cut.
Monte Carlo simulations of these effects indicate a spread in apparent
multiplicity of 5 to 6 units, at the one-sigma level, for events
with a fixed real multiplicity.
Given the low sensitivity to the $n_{ch}$ cut of the RQMD results,
the trigger differences between experiment and model are not of
consequence.

Another sample of RQMD events is also used in the comparisons to the
data.  This sample of 520 events with $0<b<2.5$ fm was calculated with
the attractive part of the quasi-potentials acting between baryon
resonances modified as explained in Ref.~5, i.e., with the delta-delta
and nucleon-baryon attractions turned off.  This sample is referred to
below as the RQMD2 sample.  No $n_{ch}$ filter was set
since more than 95\% of the events satisfied the
$n_{ch}\geq 154$ criterion.
\vglue 0.25truein
\leftline {\bf 3.~~SPECTRA AND RAPIDITY DISTRIBUTIONS}

Spectra in the form of the invariant cross section normalized per event
versus $m_t-m$ (where $m_t=\sqrt{p_t^2 + m^2}$,
$p_t= p\sin \theta$, where $p$ is the momentum and $m$ the rest mass)
were constructed from the RQMD events for
protons, $\pi^{\pm}$ and K$^{\pm}$.  Each spectrum was fit to an
exponential with the maximum likelihood procedure for Poisson statistics
within the experimental $m_t-m$ acceptance for rapidity bins of
$\Delta y=0.3$, (0.5 for K$^-$):
$${1\over N_{trig}} \bigg(E {d^3n\over dp^3}\bigg) =
A \exp(-(m_t-m)/B)~~~,\eqno	(1)$$
where $N_{trig}$=the number of events in the sample=806 for the present
RQMD set, and $A$ and $B$ are fit parameters, with $B$ called the
inverse slope parameter.  The integral of multiplicity density
(per event), $dn/dy$, was evaluated as
$$dn/dy = \int^{2\pi}_0 d\phi \int^\infty_m A \exp (-(m_t-m)/B) m_t dm_t
= 2\pi A(B^2 + mB)~~~. \eqno(2)$$
The distribution $dn/dy$ can also be found directly
by counting the number of RQMD particles in each rapidity interval.
\vglue 0.25truein
\leftline {\bf 3.1.~~Acceptance}

The RQMD hyperons and K$^0_s$ were decayed immediately after the
collision and thus no account was taken for the variation of the
experimental acceptance with the
position of the decay vertex.  In E802 only tracks that point back to
the target are accepted: this is defined by a $\pm$3~cm cut in the
$x$ and $y$ directions, where the ($x$,$y$) plane is defined as a
plane perpendicular to the beam direction ($z$-axis) with its origin
at the target center ($z$=0) (see Ref.~14.)
For the hyperons, the decay protons will largely be counted, but the
decay pions will not.  The overwhelming part of these pions will have
$m_t$ values below the E802 $m_t$ acceptance, so the effect for the
comparisons presented here is very small.  For the K$^0_s$,
it is evaluated that about 5\% of the decay pions will not be
accepted by E802, so the RQMD estimates made here
within the E802 acceptance
overcount the $\pi^{\pm}$ by less than 2\%.
\vglue 0.25truein
\leftline {\bf 3.2.~~Protons}

Figure 2 shows a comparison of RQMD proton spectra at $y=0.7$ and
$y=1.90$ to E802 spectra at the same rapidities.
(Note that the $y=1.90$ spectra have been multiplied by 0.01 for
display reasons.)  At both rapidities the RQMD spectra fall off
more steeply with $m_t$ than do the experimental spectra; i.e.,
$B_{RQMD}<B_{E802}$.  This is a general feature in the total
experimental rapidity acceptance $0.4<y<2.2$.  The $dn/dy$ as obtained
via Eqs. (1) and (2) are shown in Fig.~3, where RQMD fits in the E802
acceptance are shown as open squares.
The RQMD model does not produce clusters (d, t, $^3$He, etc.),
so it is reasonable to add the measured deuterons,
which contribute most of these clusters, to the measured
protons for the comparison.  The filled octagons show the p+d results
from E802, when p and d multiplicities are added at the same rapidity.
Statistical uncertainties are shown only when they are larger than
the size of the symbol.  The shapes of the two $dn/dy$ distributions
are different: RQMD is convex and E802 is concave.  For $0.75<y<2.0$
the RQMD points are above the experimental points.

Figure 3 also shows the E802 proton and deuteron results from a 2\%
cut on apparent multiplicity (black squares).  This cut corresponds
to the centrality cut used in Experiment E814 (Ref.~15), and
represents the most central cut reasonable within the statistics
of E802.  The RQMD points are still above the data for the
midrapidity region, $1.0 < y < 2.0$.

The slopes, $B$, of the RQMD proton spectra depend on the
quasi-potentials acting between the baryon resonances used in the
calculations (see Ref.~5.)
The calculation which produced the RQMD2 sample
(the same calculation described in Ref.~5), yields larger
inverse slope parameters, i.e., a less steep fall-off with $m_t-m$.
Figure 4 shows a comparison of the $B$-values obtained from fitting
proton spectra in the E802 experimental $m_t-m$ acceptance.  The black
squares are from the experiment, the open squares show the results from
the standard RQMD calculations (no error bars shown), while the open
octagons are from the calculations with modified potentials (RQMD2).
The standard RQMD always underpredicts the observed $B$-values,
while the RQMD2 does better at low rapidities, but also
underpredicts $B$ significantly for $y>1.2$.  Addition of the E-802
deuteron spectra to the proton spectra (at 1/2 the deuteron $m_t-m$
values) makes no change in this result.  The $dn/dy$ values
for RQMD and RQMD2 agree well with one another for $y\geq$1.0, while
RQMD2 is 10 to 20\% above the RQMD values for $y$=0.45 and 0.75.
The measured slope parameters for the 2\% and 7\% cuts agree with
one another, and the peaking of the $B$-values at $y=1.5$ seen in
Fig.~4 is repeated for the 2\% cut.

The model has too many protons in midrapidity (Fig.~3), for which
there are two possible causes: the first is too much
``stopping'', i.e., the interactions slow down the protons too much
while creating too many pions (see below.)
The second possible cause is an excessive conversion of neutrons to
protons from an overly rapid chemical equilibration, an effect that
will cause an overproduction of $\pi^-$ relative to $\pi^+$, as is
also observed (see below.)
Indeed, RQMD predicts a neutron/proton ratio of very nearly n/p=1.0
for rapidities, $y \geq 1.0$, indicating a high degree of
proton-neutron equilibration in the model.
\vglue 0.25truein
\leftline {\bf 3.3.~~Pions}

The $\pi^+$ spectra for $y=1.10$ and 2.50 are shown in Fig.~5.
The agreement between experiment (black squares) and theory
(open squares) is good.
Close examination of the RQMD pion spectra reveals that they are in
general not exponential; they often exhibit a steep rise at low
$m_t-m$ (see, e.g., the spectrum for $y=1.10$).

The spectra may be fitted to a single exponential within the
experimental $m_t-m$ acceptance with $\chi^2$ values per degree of
freedom around 1-1.5, and the slope parameters for theory and
experiment agreeing well.
The resulting $dn/dy$ values are shown in Fig.~6a, for both $\pi^+$
and $\pi^-$ and with the $\pi^-$ values multiplied by 0.1 for
display purposes.  The RQMD $dn/dy$ are generally higher than the
experimental values by $\approx$10\% for $\pi^+$ and $\approx$15\%
for $\pi^-$.  The distribution shapes are very similar.

The $\pi^+$ $dn/dy$ values for the 2\%  multiplicity
cut (not shown) give quantitative agreement with  RQMD for
$1.0\leq y \leq 2.0$, while there is still a small
discrepancy at the outer ranges of the rapidity interval.  RQMD2 gives
a pion yield close to RQMD.  The model acts as a somewhat more central
event sample than the 7\% cross section cut would indicate, a trend
that is not inconsistent with the proton discussion above.

The RQMD pion spectra in the full $m_t-m$ acceptance require a sum
of two exponentials to be adequately fit, where one component has
a $B$ value near 80 MeV and the other has $B\approx$ 150 MeV.
In cases like this, the total $dn/dy$ evaluated from
a fit in the experimental acceptance and from counting in the full
acceptance are of course quite
different, as demonstrated in Fig.~6b for $\pi^+$.
The RQMD model shows that the steep low-$p_t$ part of the pion spectra
(which is outside the E802 acceptance) originates predominantly from
decays of $\Delta$-resonances$^{8,16}$.  ($\Lambda$ decays do not
contribute to the $\pi^+$ spectrum, but are important for $\pi^-$.)
The low $p_t$ rise for pions, predicted by RQMD, has been observed in
AGS experiments E810$^{17}$ and E814$^{18}$.
\vglue 0.25truein
\leftline {\bf 3.4.~~Kaons}

The agreement between data and theory for kaons is of similar
quality as for pions.  This is reflected in the $dn/dy$
integrals shown in Fig.~7 for both K$^+$ and K$^-$.  On the average
the E802 K$^+$ $dn/dy$ are $\approx$20\% higher than the RQMD values,
while the K$^-$ measurements are $\approx$20\% lower than the theory.
The distribution shapes agree well.
The experimental values of the slope parameter, $B$ for K$^+$ are
larger than the RQMD predictions, typically $\approx$220 MeV vs.
$\approx$180 MeV.

The conventional ratios K$^+/\pi^+$ and K$^+/{\rm K}^-$ are
displayed in Fig.~8 versus rapidity for both E802 and RQMD.  The
experimental K$^+/\pi^+$ ratio is consistently higher than the RQMD
ratio, an effect that is not due to the low-p$_t$ rise in the RQMD
$\pi^+$ spectra, as all fits were made in the E802 acceptance.
Also the K$^+/{\rm K}^-$ ratios are different,
with the E802 ratio being consistently higher.  The 2\% centrality cut
for E802 does not change any of the ratios discussed, so it is unlikely
that the discrepancy is caused by differences in the trigger conditions.
In the RQMD model there are substantial contributions to the K$^+$ and
K$^-$ production from baryon-meson interactions and some contribution to
K$^-$ from meson-meson interactions$^6$.  These cascade-like
interactions depend strongly on the dynamics of the nuclear collisions,
and it is unlikely that a somewhat schematic model such as RQMD should
describe this complex situation entirely accurately.  The deviations
\underbar {are} emphasized in the ratios shown.
\bigskip
\leftline {\bf 4.~~ AVERAGE MULTIPLICITIES AND MOMENTUM SUMS}

Table 1 shows the multiplicities for protons, pions and kaons summed
over the E802 rapidity acceptance. The deuteron multiplicities have
been added to the proton multiplicities for E802. The ratios of
E802/RQMD multiplicities in the last column show that the RQMD model
overpredicts the particle yield by about 25\% on the average. For
K$^+$ the model underpredicts by a similar amount.

Both $\pi^+$ and $\pi^-$ have been measured in identical rapidity
intervals, so the particle ratio
$R(\pi) = \Delta n(\pi^-)/\Delta n(\pi^+)$ is of relevance.
Because of the neutron surplus in the Au target, it is expected that
the ratio should be larger than unity.
Table 1 yields $R(\pi)$=1.09$\pm$0.02 for E802 and 1.15$\pm$0.01 for
RQMD.  The difference in the E802 and RQMD ratios is near three
standard deviations and may be significant. (See also the discussion
in Sect. 3.3.)

The total transverse momentum production shown in Table~2 was
calculated from the exponential fits (Eq.~1) in the E802 $m_t-m$
interval by:
$$P_t = \sum^{y_2}_{y_1}\langle p_t \rangle \bigg( {dn\over dy} \bigg)
\Delta y~~~, \eqno (3)$$
where $y_1$ and $y_2$ are the rapidity limits given in the table,
$\Delta y$ is the bin size, and
$$\langle p_t \rangle=
{1\over dn/dy}\int^{2\pi}_0 d\phi \int^\infty_m A\exp(-(m_t-m)/B)~
m_t\sqrt{m_t^2-m^2}~dm_t \eqno (4)$$
$$= {m^2\over B+m}~\exp(m/B)~K_2(m/B)~~~,$$
where $K_2$ is a modified Bessel function of second order.

The total longitudinal momemtum, $P_{\parallel}$,
is defined as
  $$P_{\parallel}=\sum^{y_2}_{y_1} \langle p_{\parallel} \rangle
\bigg( {dn\over dy}\bigg) \Delta y  \eqno    (5)$$
with
  $$\langle p_{\parallel} \rangle
=\langle m_t \rangle \sinh y={1\over dn/dy}~\int_0^{2\pi}d\phi
\int_m^{\infty} A\exp(-(m_t-m)/B)~m_t^2~\sinh y~dm_t$$
$$=\sinh y\cdot 2B\cdot {1+m/B+m^2/2B^2\over 1+m/B}~~~.   \eqno (6)$$

For protons (with deuterons added for E802), the $P_t$ ratio E802/RQMD
is 0.96, compared with 0.84 for $P_{\parallel}$ and 0.73 for the
multiplicities, $\Delta n$.  The experiment clearly has larger
$\langle p_t \rangle$ and $\langle p_{\parallel} \rangle$ values than
RQMD.  This is reversed for the pions, for which the E802/RQMD ratio
is smaller for the momenta than for the multiplicities, although the
effect is relatively smaller than for the protons.  The K$^+$ trend is
like that for the protons.
This analysis supports the discussion given in the above sections.

$P_{\parallel}$ is a conserved quantity, so the less than unity ratio in
Table 2 for ``All'' indicates that the E802/RQMD ratio at rapidities
higher than those covered by the E802 acceptance --- higher rapidities
rather than lower because of the $\sinh(y)$ weighting in Eq.~6 --- should
be going from below one to above one.
This conclusion is similar to the one drawn from other model comparisons in
Ref.~19, except that the E802 deficit is much smaller in the present
comparison.

\vfil\eject
\parskip=12pt
\parindent 0.5truein
\leftline {\bf 5~~CONCLUSIONS}

The overall agreement between the RQMD model and the E802
experimental data is quite satisfactory, mostly within $\approx$20\%.
(Recall that there is a 10-15\% systematic
error associated with the experimental results$^{14}$.)
The largest differences between RQMD and E802 occur for the proton
$dn/dy$ distributions (Fig.~3) and for the proton inverse
slope parameter distributions (Fig.~4).

It is our conclusion that the calculation gives too many
protons in the central rapidity region as compared to the
E802 data.
The relevant processes for this disagreement are a difference in the
baryon stopping power, and too strong of a removal of the initial
target neutron excess.
A contribution from the latter process receives support from the
excess of negative mesons over positive mesons predicted by the
model and not observed experimentally; also predicted is a
neutron/proton ratio near 1.0 for $y \geq 1.0$.

The further discrepency, the higher average $p_t$ around
midrapidity, is just the reflection of the larger
experimental proton slope parameters as discussed above.
While the density dependence of the quasipotentials can
be successfully manipulated to harden the proton slopes
in the center of the participant fireball ($y \in (1,1.2)$),
it does not give sufficient repulsion for the protons
forward of rapidity 1.2.  This may point towards an
additional repulsion due to the momentum dependence of
nuclear mean fields in dense matter.  It is well known
that such a momentum dependence is present at ground-state
density which is experimentally accessible via optical
potential measurements in p+$A$ reactions.  However, no
safe knowledge exists on how the density and momentum
dependence of mean fields are intertwined.

The large experimentally observed splitting between the proton
and the pion slope parameters --- which is qualitatively
confirmed by RQMD and even somewhat underestimated in the
forward rapidity region --- is  rather remarkable,
and may point towards the importance of collective flow
in nucleus-nucleus reactions at AGS energy.
The collective flow component of the transverse momenta
increases with the mass of a particle.  In RQMD this is the
most important mechanism which leads to a slope parameter
splitting for the various hadron species.

In general, the RQMD model seems to describe reasonably well the
mechanism of heavy ion reactions at AGS energies.
The hadron distributions from central
nucleus-nucleus collisions can not be explained by simple
superpositions of nucleon-nucleon collisions.
The comparison of the data with the predictions of the
transport model shows
clearly that the rescattering of produced particles in the hadronic
matter plays a major role in the evolution of the collisions.
In particular, the production and rescattering of hadron resonances
are the key processes to understanding the stopping power, transverse
mass spectra and strangeness enhancement observed experimentally.
\vglue 0.25 truein
\leftline {\bf ACKNOWLEDGEMENTS}

This work has been supported in part by the U.S. Department of Energy
under contracts with Brookhaven National Laboratory and Los Alamos
National Laboratory.  R. Mattiello wishes to express his gratitude
to Drs. W. Greiner and H. St\"{o}cker for providing excellent working
conditions and for many helpful discussions.
\vfil\eject
\parskip=12pt
\parindent 0.5truein
\leftline {\bf REFERENCES}
\item {a)} Now at LPNHE-Ecole Polytechnique, F-91178 Palaiseau, France.
\item {b)} Now at the Niels Bohr Institute for Astronomy, Physics and
Geophysics, DK-2100 Copenhagen \O, Denmark.
\item {c)} Now at the
Institut f\"ur Theoretische Physik, Johann Wolfgang Goethe
Universit\"at, Frankfurt am Main, Germany.
\item {1)} B. Nilsson-Almqvist and E. Stenlund, Computer Phys. Comm.
{\bf 43}, 387 (1987); B. Andersson, G. Gustafson and B.
Nilsson-Almqvist, Nucl. Phys. {\bf B281}, 289 (1987).
\item {2)} H. Sorge, H. St\"ocker and W. Greiner, Ann. Phys. (NY)
{\bf 192}, 266 (1989); Nucl. Phys. {\bf A498}, 567c (1989); Z.
Phys. {\bf C47}, 629 (1990).
\item {3)} K. Werner, Z. Phys. {\bf C42}, 85 (1989).
\item {4)} Y. Pang, T.J. Schlagel and S. Kahana, Phys. Rev. Lett.
{\bf 68}, 2743 (1992).
\item {5)} H. Sorge, R. Mattiello, H. St\"ocker and W. Greiner, Phys.
Rev. Lett. {\bf 68}, 286 (1992).
\item {6)} R. Mattiello, H. Sorge, H. St\"ocker and W. Greiner, Phys.
Rev. Lett. {\bf 63}, 1459 (1989).
\item {7)} H. Sorge, A. von Keitz, R. Mattiello, H. St\"ocker and W.
Greiner, Phys. Lett. {\bf B243}, 7 (1990).
\item {8)} H. Sorge, R. Mattiello, A. Jahns and W. Greiner, Phys. Lett.
{\bf B271}, 37 (1991).
\item {9)} A. Jahns, H. St\"ocker, W. Greiner and H. Sorge, Phys. Rev.
Lett. {\bf 68}, 2895 (1992).
\item {10)} T. Abbott {\it et al.}, The E802 Collaboration, Nucl.
Instrum. and Methods {\bf A290}, 41 (1990).
\item {11)} T. Abbott {\it et al.}, The E802 Collaboration, Phys. Rev.
Lett. {\bf 64}, 847 (1990).
\item {12)} T. Abbott {\it et al.}, The E802 Collaboration, Phys. Rev.
Lett. {\bf 66}, 1567 (1991).
\item {13)} J.V. Allaby {\it et al.}, CERN Report No. 70-12, 1970
(unpublished); H. B\o ggild {\it et al.}, Nucl. Phys. {\bf B57},
77 (1973);
D. Dekkers {\it et al.}, Phys. Rev. {\bf 137}, B962 (1965); U. Becker
{\it et al.}, Phys. Rev. Lett. {\bf 37}, 1731 (1976).
\item {14)} T. Abbott {\it et al.}, The E802 Collaboration,
Phys. Rev. {\bf C50}, 1024 (1994).
\item {15)} P. Braun-Munzinger, The E814 Collaboration,
Nucl. Phys. {\bf A544}, 137c (1992).
\item {16)} H. Sorge, Phys. Rev. {\bf C49}, R1253 (1994).
\item {17)} S. Ahmad {\it et al.}, The E810 Collaboration, Phys. Lett.
{\bf B281}, 29 (1992).
\item {18)} T. Hemmick, The E814 Collaboration, Nucl. Phys.
{\bf A566}, 435c (1994).
\item {19)} S. Chapman and M. Gyulassy, Phys. Rev. Lett.
{\bf 67}, 1210 (1991).
\vfil\eject
\tabskip 0.40truein
$$\halign {\hfil#\hfil&\hfil#\hfil&\hfil#\hfil&\hfil#
\hfil&\hfil#\hfil&\hfil#\hfil&\hfil#\hfil&\hfil#\hfil&\hfil#\hfil&\hfil#\hfil
&\hfil#\hfil&\hfil#\hfil&\qquad\hfil#\hfil\cr
\noalign{                    \centerline {Table 1}}
\noalign{                    \centerline {Average multiplicities}}
Particle &Rapidities &$\Delta n$(RQMD) &$\Delta n$(E802) & E802/RQMD\cr
 p (+d)  & 0.4-2.2   & 45.2$\pm$0.2    &33.0$\pm$0.2     & 0.73$\pm$0.01\cr
 $\pi^+$ & 0.6-2.8   & 32.4$\pm$0.2    &27.0$\pm$0.3     & 0.83$\pm$0.01\cr
 $\pi^-$ & 0.6-2.8   & 37.3$\pm$0.3    &29.5$\pm$0.3	 & 0.79$\pm$0.01\cr
 K$^+$   & 0.6-2.2   & 3.5$\pm$0.1     & 4.3$\pm$0.1     & 1.22$\pm$0.05\cr
 K$^-$   & 0.7-2.3   & 1.03$\pm$0.04   & 0.90$\pm$0.06   & 0.87$\pm$0.07\cr
 All     &           & 119.4$\pm$0.4   &94.7$\pm$0.5     & 0.79$\pm$0.01\cr
}$$
\vglue 1.0truein
\parskip 0pt
\parindent 0.0truein
\tabskip 0.20truein
$$\halign {\hfil#\hfil&\hfil#\hfil&\hfil#\hfil&\hfil#
\hfil&\hfil#\hfil&\hfil#\hfil&\hfil#\hfil&\hfil#\hfil&\hfil#\hfil&\hfil#\hfil
&\hfil#\hfil&\hfil#\hfil&\qquad\hfil#\hfil\cr
\noalign{\centerline{Table 2}}
\noalign {\centerline{Momentum Sums}}
\noalign {\quad Part.~~Rapidities\hglue 0.8truein $P_t$\hglue
1.8truein$P_{\parallel}$\hglue 1.2truein E802/RQMD}
\noalign {\hglue 2.0truein GeV/$c$\hglue 1.5truein GeV/$c$}
&&E802&RQMD&E802&RQMD&$P_t$&$P_\parallel$\cr
p&0.4-2.2&23.7$\pm$0.2&26.1$\pm$0.2&63.1$\pm$0.3&78.3$\pm$0.6&0.91$\pm$0.01&
0.81$\pm$0.01\cr
d&0.4-1.6&2.6$\pm$0.1&&5.7$\pm$0.1&&&&\cr
p+d&&&&&&0.96$\pm$0.01*&0.84$\pm$0.01*\cr
$\pi^+$&0.6-2.8&9.2$\pm$0.1&11.8$\pm$0.1&26.4$\pm$0.2&33.8$\pm$0.3&
0.78$\pm$0.01&0.78$\pm$0.01\cr
$\pi^-$&0.6-2.8&9.9$\pm$0.1&13.5$\pm$0.1&28.1$\pm$0.2&38.3$\pm$0.4&
0.73$\pm$0.01&0.73$\pm$0.01\cr
K$^+$&0.6-2.2&2.3$\pm$0.1&1.63$\pm$0.05&6.4$\pm$0.2&4.9$\pm$0.1&
1.4$\pm$0.1&1.3$\pm$0.1\cr
K$^-$&0.7-2.3&0.46$\pm$0.05&0.52$\pm$0.03&1.2$\pm$0.1&1.7$\pm$0.1&
0.88$\pm$0.11&0.71$\pm$0.07\cr
All*&&46.9$\pm$0.3&53.6$\pm$0.3&128.1$\pm$0.5&157.0$\pm$0.8&0.88$\pm$0.01&
0.82$\pm$0.05\cr
}$$
$^*$ Deuterons added to the E802 value with 50\% of the quoted momentum
value.
\vfil\eject
\parskip=12pt
\parindent 0.5truein
\leftline {\bf Figure Captions}
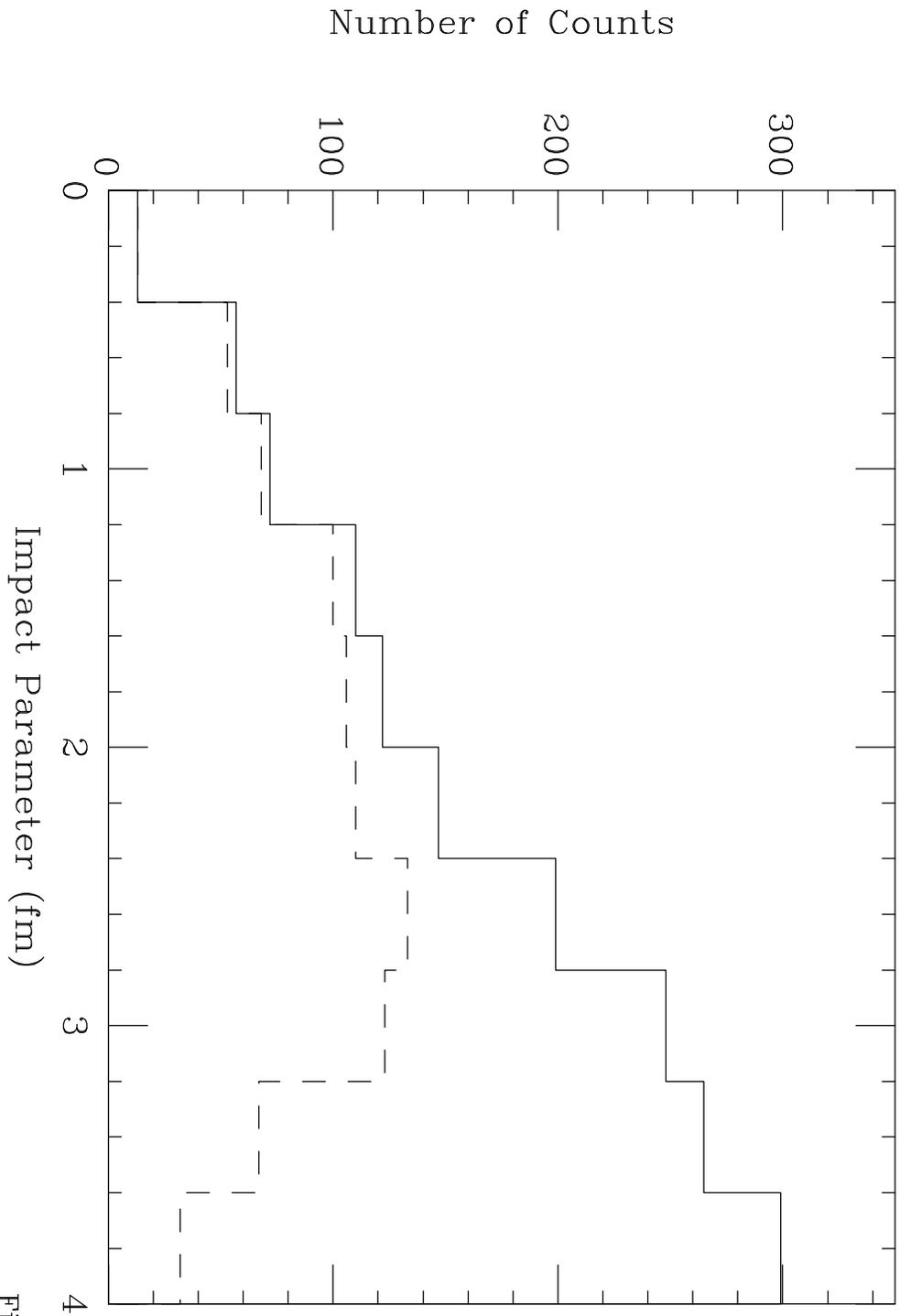
\item {Fig. 1}
The impact parameter distributions for the entire RQMD event sample
(1537 events, fully drawn histogram) and the ``central'' sample
(806 events, broken histogram).
The central distribution was obtained by setting conditions on proton
and pion momenta and the charged multiplicity in the acceptance of
the E802 multiplicity detector (see the text for details).

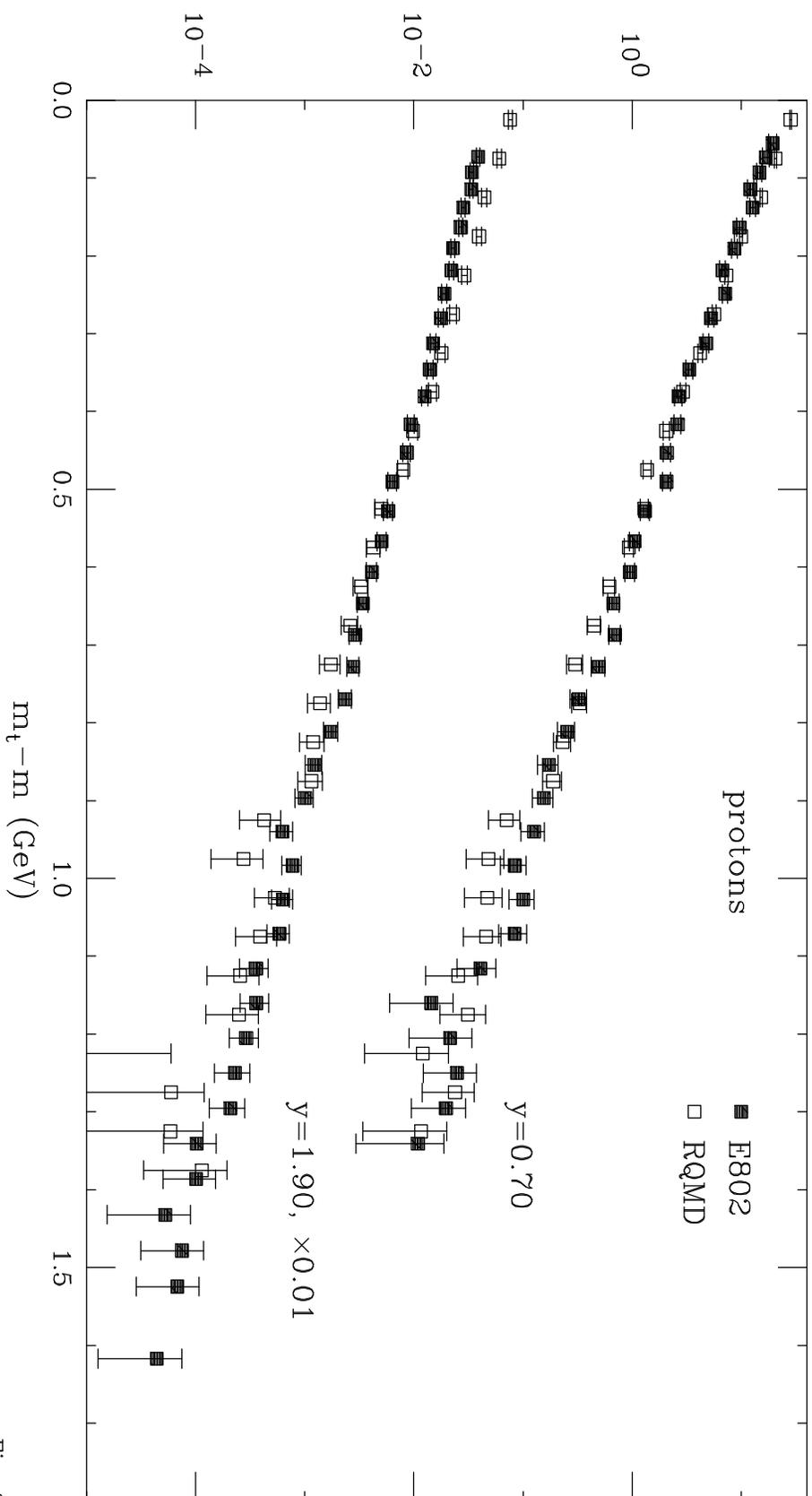
\item {Fig. 2}
Comparison of proton spectra from E802 (filled squares)
and RQMD (open squares).
The invariant cross-section per event is plotted versus transverse
kinetic energy $m_t-m$, for two different rapidity intervals
$y$=0.70 and $y$=1.90.  For E802 the interval width is
$\Delta y$=0.2, while it is 0.3 for RQMD.
The error bars are statistical, assuming a Poisson distribution.
The $y$=1.90 cross sections have been multiplied by 0.01 for
reasons of display.

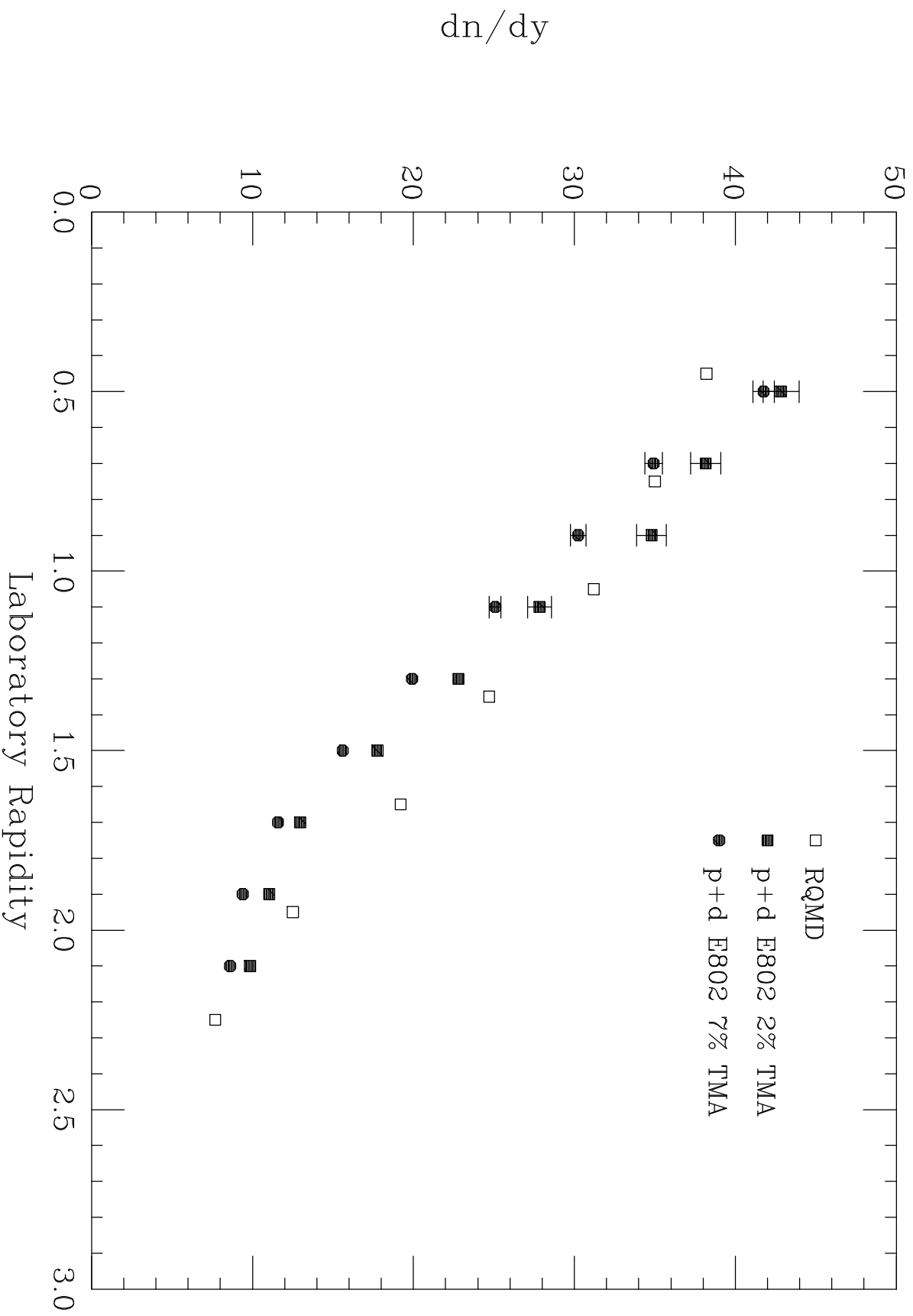
\item {Fig. 3}
Proton rapidity distributions $dn/dy$.
The open squares denote the multiplicities per unit rapidity
obtained by integration of the RQMD proton spectra within the E802
$m_t$ acceptance.
Black octagons denote the E802 $dn/dy$, where the deuteron yield has
been added to the proton $dn/dy$ at the same rapidity.
Black squares are $p+d$ E802 $dn/dy$ for a more central multiplicity
cut, 2\% of $\sigma_{inel}$ instead of the standard 7\%.

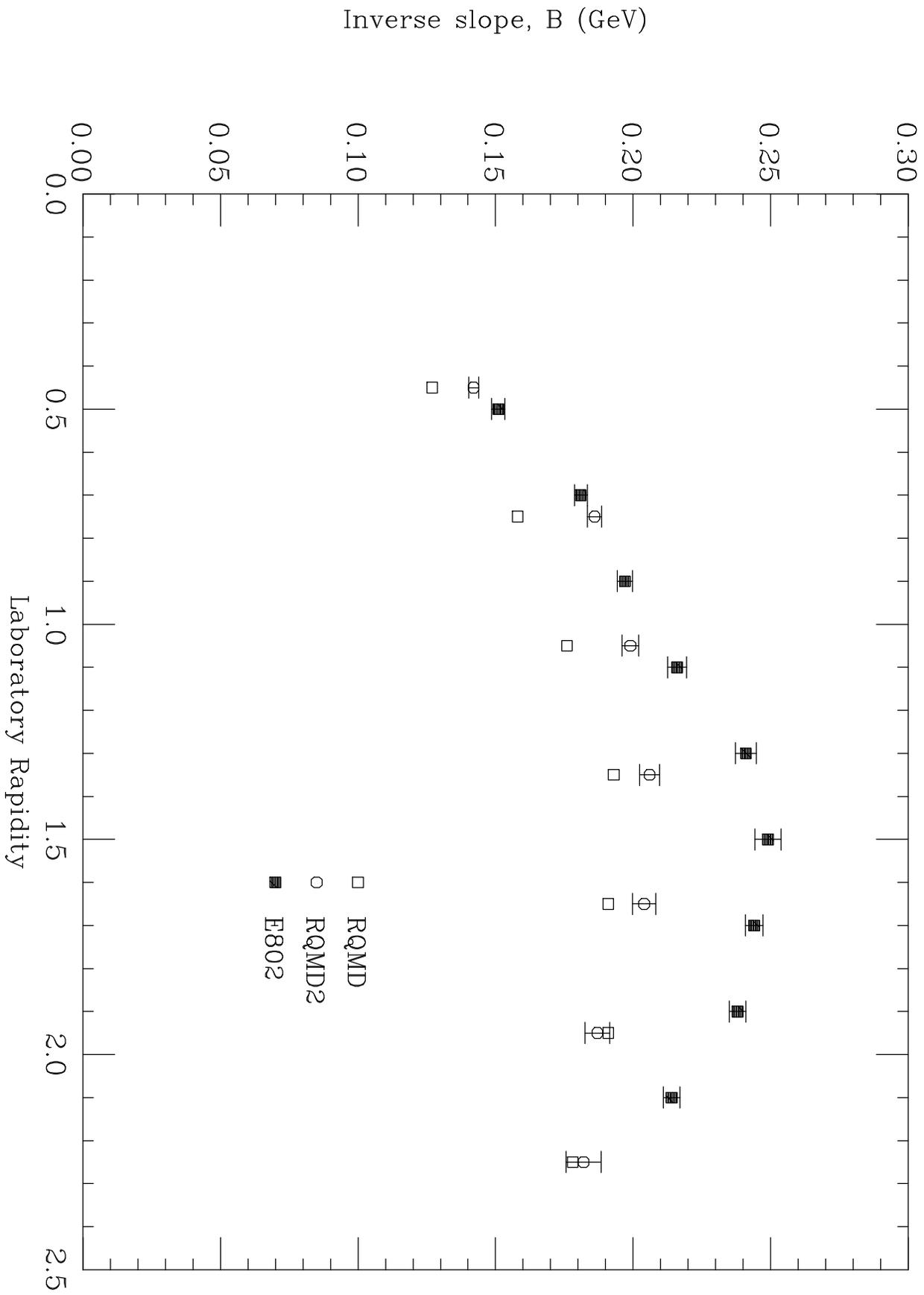
\item {Fig. 4}
Inverse slope parameters $B$ (in GeV), as defined in Eq. (1),
plotted versus rapidity for protons.
Black squares denote E802 data, open squares RQMD results in the E802
acceptance and finally open octagons stand for RQMD2 results in the
experimental acceptance.

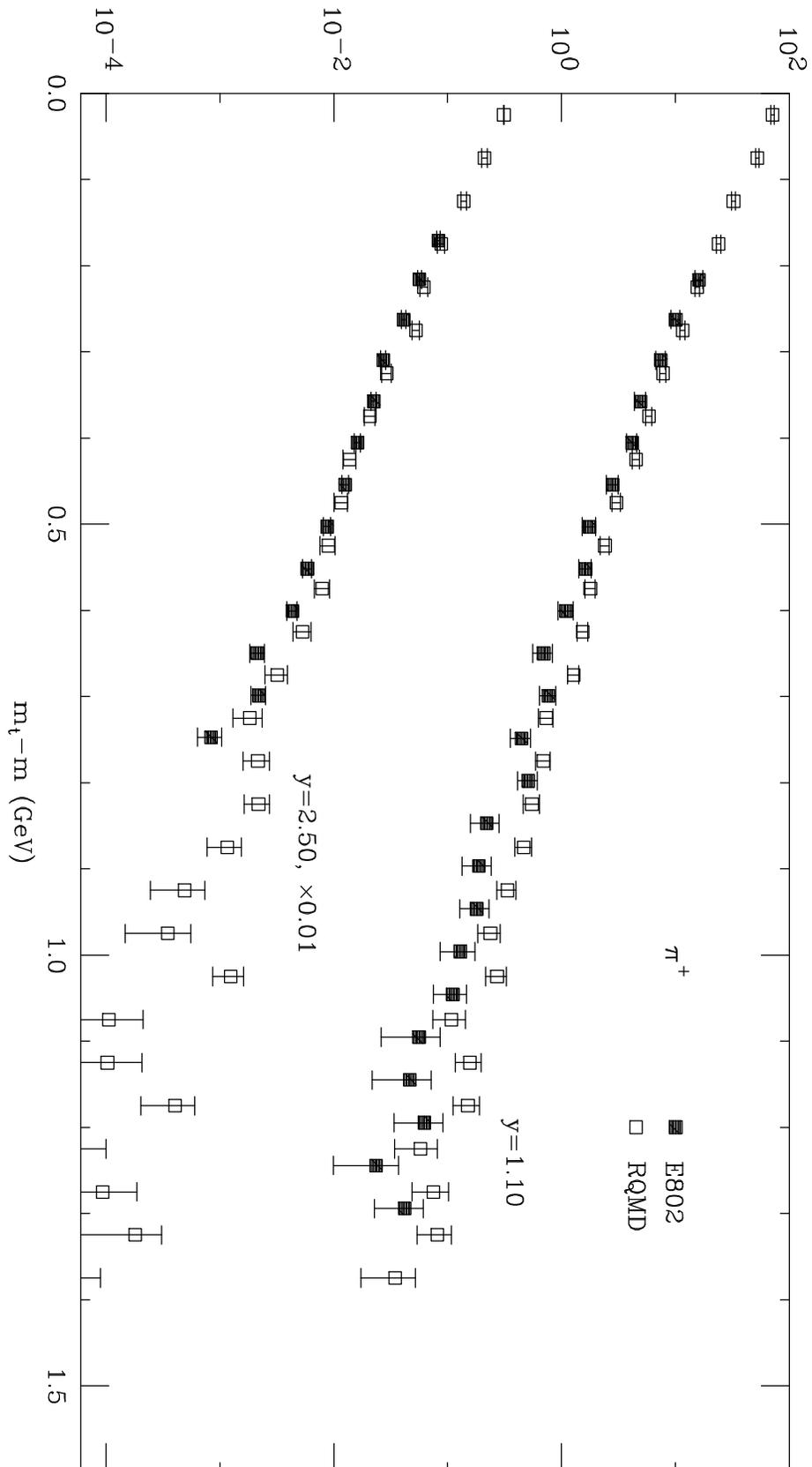
\item {Fig. 5}
Spectra for $\pi^{+}$ in the two rapidity intervals $y$=1.10 and
$y$=2.50.
Open squares are from RQMD, while black squares are from E802.
The $y$=2.50 invariant cross sections have been multiplied by 0.01
for reasons of display.
The rapidity range is $\Delta y$=0.2 for E802 and 0.3 for RQMD.
Error bars are statistical only.

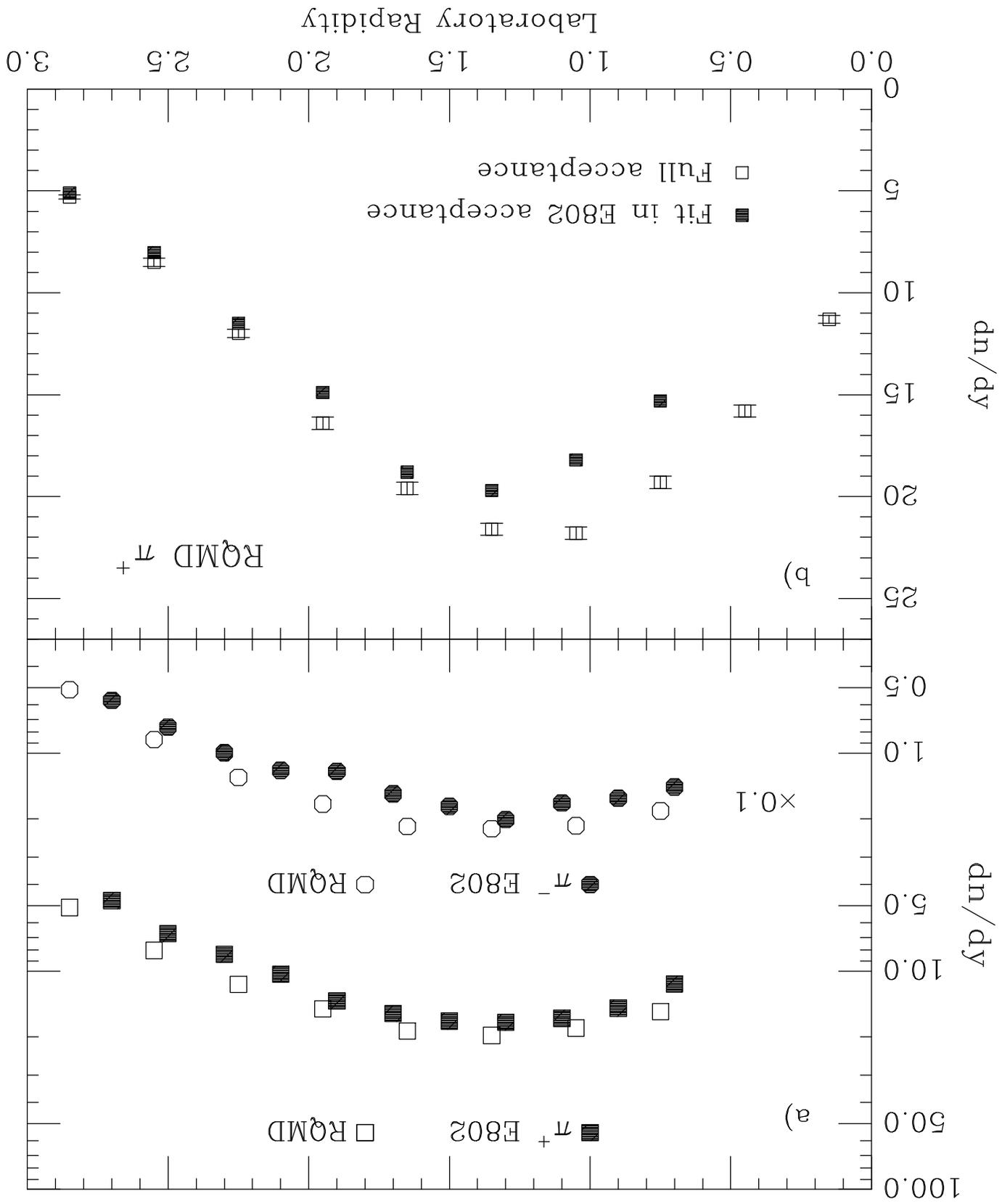
\item {Fig. 6}
a) Rapidity distributions $dn/dy$ for $\pi^+$ (squares) and
$\pi^-$ (octagons) from E802 (black symbols) and RQMD analyzed
in the experimental acceptance (open symbols).  The $\pi^-$
$dn/dy$ have been divided by 10 for reasons of display.
b) Comparison between $dn/dy$ for $\pi^+$ from RQMD from
in two different $m_t-m$ acceptances.
The black points are from single exponential fits in the E802
acceptance, while the open points were obtained by counting over
the entire $m_t-m$ range.

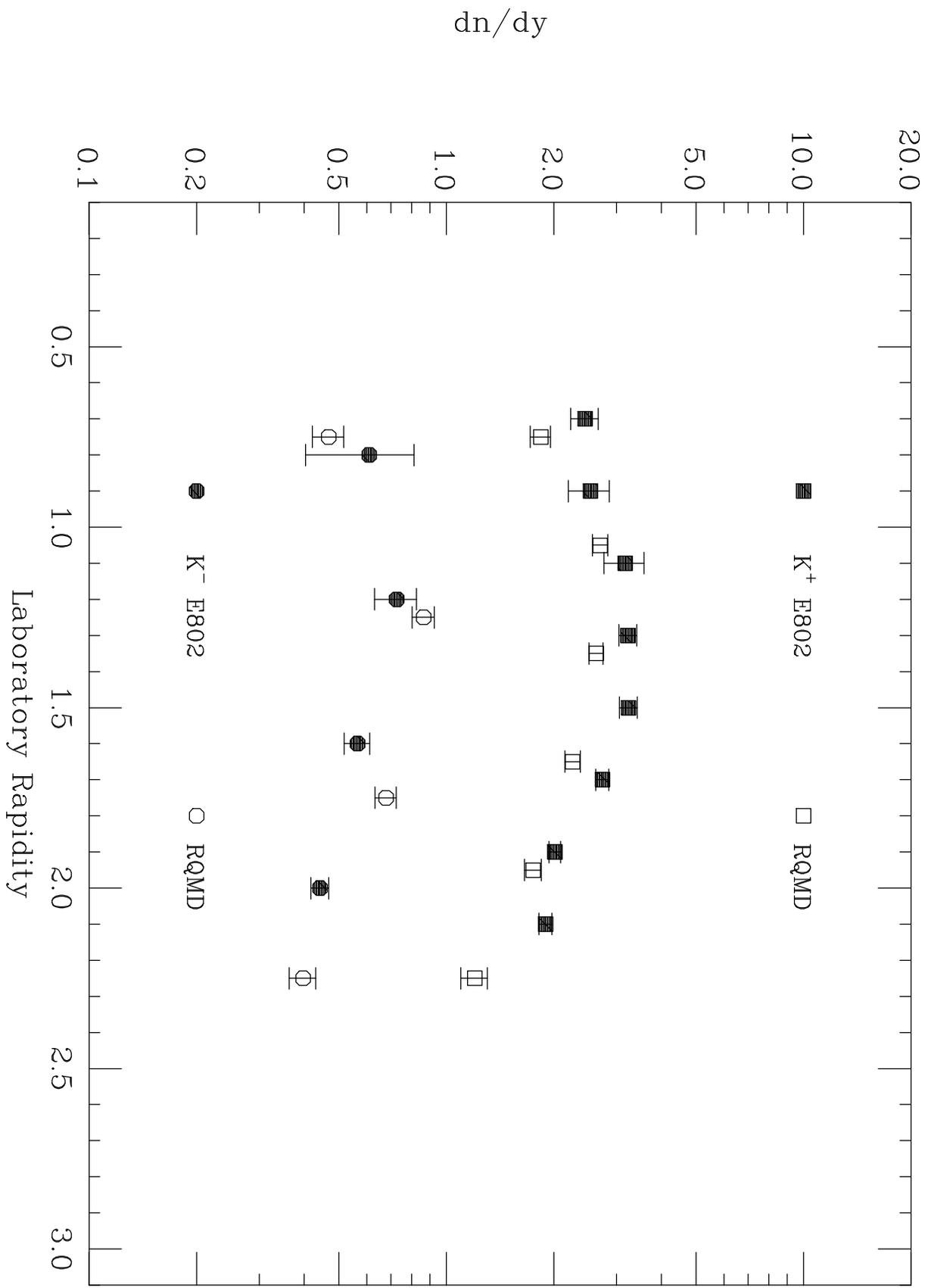
\item {Fig. 7}
Rapidity distributions $dn/dy$ for K$^+$ (squares) and
K$^-$ (octagons) from E802 (black symbols) and RQMD
analyzed in the experimental acceptance (open symbols).

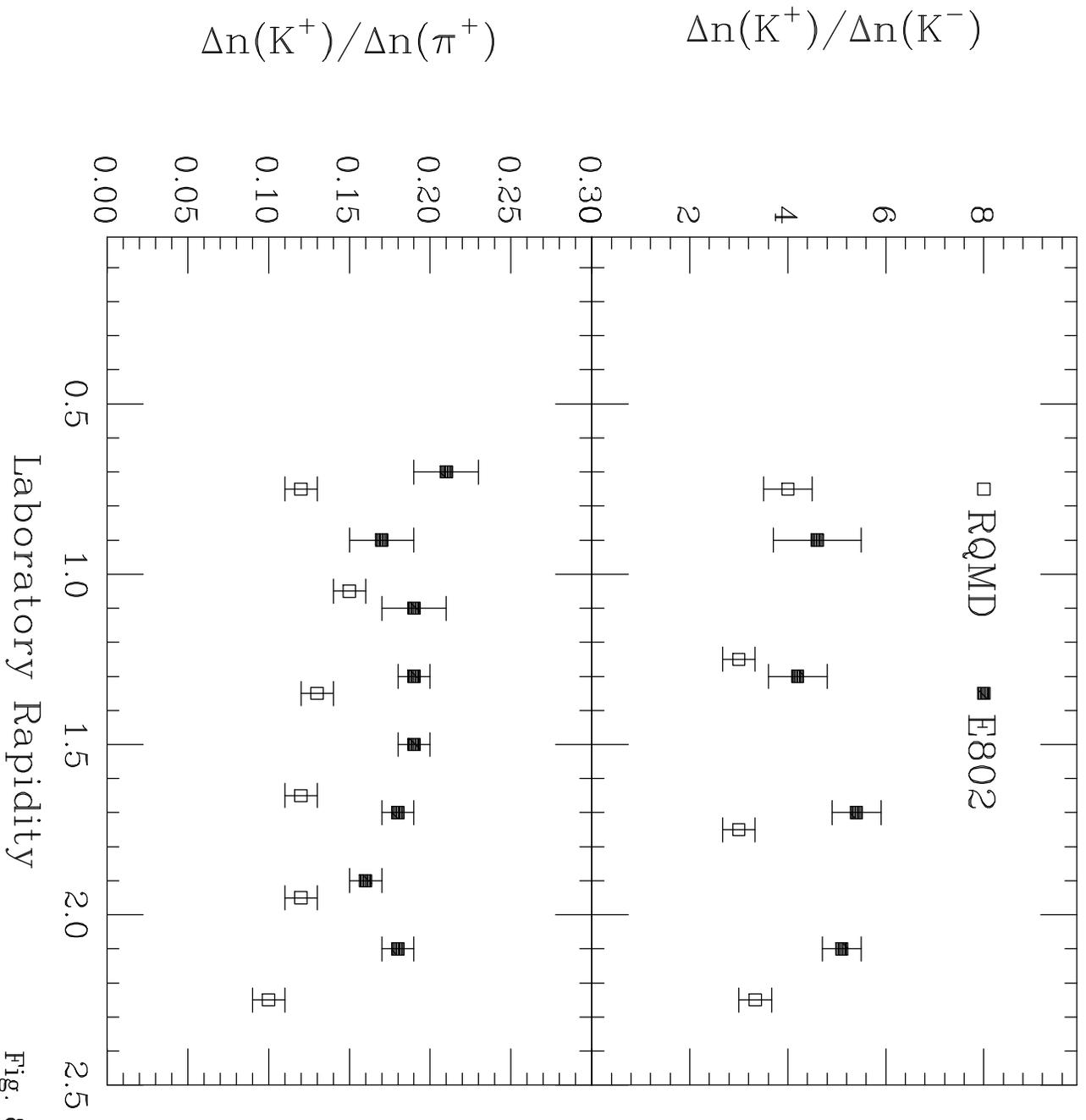
\item {Fig. 8}
Comparison of multiplicity ratios from E802 (black symbols) and RQMD
analyzed in the experimental acceptance (open symbols).
The upper frame gives the $\Delta n($K$^+)/\Delta n($K$^-)$ ratio
plotted against laboratory rapidity, while the
lower frame shows the ratio $\Delta n($K$^+)/\Delta n(\pi^+)$.

\vfil\eject
\end